**Isotopically Selected Single Antimony Molecule Doping**

Mason Adshead[1*], Maddison Coke[1*], Evan Tillotson[2], Kexue Li[2], Sam Sullivan-Allsop[2], Ricardo Egoavil[3], William Thornley[2], Yi Cui[4], Christopher M. Gourlay[4], Katie L Moore[2], Sarah J Haigh[2], Richard J Curry[1]

[1]Department of Electrical and Electronic Engineering, Photon Science Institute, University of Manchester, Oxford Road, Manchester M13 9PL, UK

[2]Department of Materials, Photon Science Institute, University of Manchester, Oxford Road, Manchester M13 9PL, UK

[3]Thermo Fisher Scientific, Achtseweg Noord 5, Bldg. III 3-367, 5651 GG Eindhoven, The Netherlands

[4]Department of Materials, Imperial College London, London SW7 2AZ, UK

*These authors contributed equally to this work.

**Abstract**

A reliable route to the deterministic fabrication of impurity ion donors in silicon is required to advance quantum computing architectures based upon such systems. This paper reports the ability to dope isotopically-defined unique ($^{121}$Sb$^{123}$Sb) clusters into silicon with measured detection efficiencies of 94% being obtained. Atomically resolved imaging of the doped clusters reveals a Sb-to-Sb separation of ~2 nm post-implantation, thus indicating suitability to form coupled qudit systems. The method used is fully compatible with integration into processing that includes pre-enrichment of the silicon host to < 3ppm $^{29}$Si levels. As such, we present a potential pathway to the creation of scaled qudit arrays within silicon platforms for quantum computing.

**Introduction**

The development of a quantum computer that fully realises the potential of quantum-based operations requires a scalable materials platform to facilitate its fabrication. This presents a significant challenge as an error-corrected machine will require $10^4$ to $10^6$ physical qubits, depending on their performance and operation. A variety of approaches are being explored to deliver this based upon various technology platforms, including trapped ion [1-3] superconducting [4-6], optical [7] and semiconductor [8-10] based qubit systems. Each approach has its own advantages and challenges [11, 12], but in terms of scaling to large qubit numbers, semiconductor platforms are attractive due to opportunities to leverage the mature state of microelectronics processing.

Significant effort has therefore been expended in developing Si-based architectures for quantum computing with electrostatically gate-defined quantum dots and impurity ion donors, both yielding qubit devices [13]. The former, utilising a large (232) ensemble of 12 Si:SiGe quantum dot arrays, have demonstrated average coherence times ($T_2^*$) of 0.6 μs in natural Si ($^{nat}$Si). This was increased by an order of magnitude through reducing the residual $^{29}$Si isotopic population from that of $^{nat}$Si (~46,000 ppm) to ~800 ppm, which reduces nuclear spin-related decoherence. The use of impurity ion donors, proposed by Kane [14], has delivered electron coherence values of $T_2^*$ ~55 ns in $^{31}$P-doped $^{nat}$Si, again limited by the residual $^{29}$Si nuclear spin interactions [15]. Harnessing of the nuclear spin (*I*) of such donors has also been demonstrated with values of $T_2^*$ ~

0.84 ms and 3 ms for the neutral and charged donors ($D^0$ and $D^+$) respectively, where $I$ = 1/2 for $^{31}$P [16]. These coherence times are also limited by the presence of $^{29}$Si hence efforts to produce increasingly isotopically pure $^{28}$Si have been pursued, with record enrichment levels of <3 ppm $^{29}$Si being recently reported through focused ion beam enrichment [17].

The utilization of donor ions enables the combined use of both the electron and nuclear spin to access qubits operating in a higher dimensional Hilbert space. This scales with the nuclear spin ($I$) of the donor which for $^{31}$P provides a 4-dimensional system when coupled to the electron spin ($S$ = 1/2). The utilization of higher nuclear spin donors such as $^{123}$Sb ($I$ = 7/2) has been demonstrated to enable access to qudits operating in 16-dimensional space [18]. A single such qudit has recently been used to demonstrate 'Schrodinger's cat' entangled superposition states [19]. The utilisation of qudits therefore provides a powerful opportunity for inward scaling of qubits which can then be combined with physical scaling of such systems. The gain of using a $d$-dimensional qudit over a 2-dimensional qubit is a reduction by $\log_2 d$ in the number of physical units required, a factor of 4 for $^{123}$Sb. Furthermore, coupling together two such donor systems yields a $(\log_2 d)^2$ improvement.

The potential for such scaling presents challenges and opportunities. If successfully implemented it could significantly reduce the architecture footprint of future quantum devices whilst accelerating performance. However, creating coupled qudits requires the ability to reproducibly manufacture them with high precision. Here we demonstrate the use of focused ion beam (FIB) implantation to dope Si with $Sb_2$ molecules and achieve reproducible Sb-Sb separation of sub-3 nm. We use NanoSIMS characterisation to confirm the isotopic nature of the implanted $Sb_2$ molecules and atomic resolution scanning transmission electron microscopy (STEM) to measure their separation.

**Results and Discussion**

Initial observation of the potential to isolate $Sb_2$ molecules within a FIB was first noted by Adshead *et al.* [20]. To our knowledge, the isolation and verification of such a cluster has not previously been achieved, due to a combination of reasons, which include low mass resolution instrumentation [21]. Indeed, it was not until MC-ICP-MS (Multicollector Inductively Coupled Plasma Mass Spectrometry) was available that the relative abundance of the two naturally occurring Sb isotopes ($^{121}$Sb and $^{123}$Sb) could be verified [22]. Here, the Platform for Nanoscale Advanced Materials Engineering (P-NAME) FIB Facility (Q-One, Ionoptika) is used at a 15 kV anode voltage with a liquid metal $Au_xSi_ySb_z$ eutectic alloy ion source. Figure 1 demonstrates the fully resolved Sb peaks (full Wein Filter (WF) scan is shown in SI Figure S1). The high mass resolving power of this instrument at this mass-energy range (R = m/$\Delta$m = 296 ±7) enables the identification of an additional third peak lying between the $^{121}$Sb and $^{123}$Sb peaks. This additional third peak exhibits a significantly lower concentration, and has no potential isobars which may be identified to account for its presence. It is therefore proposed that the peak is due to the detection of a ($^{121}$Sb$^{123}$Sb)$^{2+}$ molecular ion. This would also imply that associated monoisotopic $^{121}$Sb$_2^{2+}$ and $^{123}$Sb$_2^{2+}$ clusters are present within the equivalent isotopic singly-charged ion (Sb$^+$) peaks, which cannot be separated. Using the natural abundance of the Sb isotopes [23], integration of the peaks in Figure 1 yields ratios for the proposed clusters of 18.2 : 48.4: 32.7, respectively, for $^{123}$Sb$_2^{2+}$: ($^{121}$Sb$^{123}$Sb)$^{2+}$:$^{121}$Sb$_2^{2+}$, these are modelled in Figure 1 as coloured solid lines. The presence of $^{121}$Sb$_2^{2+}$ and $^{123}$Sb$_2^{2+}$ clusters has implications for attempting single Sb ion implantation using singly-charged isotopes ($^{121}$Sb$^+$ and $^{123}$Sb$^+$) as these are indistinguishable from the doubly-charged cluster Sb pairs of the same isotope, which would be present with a 6-8% probability.

Therefore, for applications requiring *single*-ion Sb qudit doping isotopically selected $Sb^{2+}$ ions should be used where there is no evidence of an associated isobar cluster with the same mass to charge ratio (e.g. $^{123}Sb_2^{4+}$) as shown in SI Figure S1.

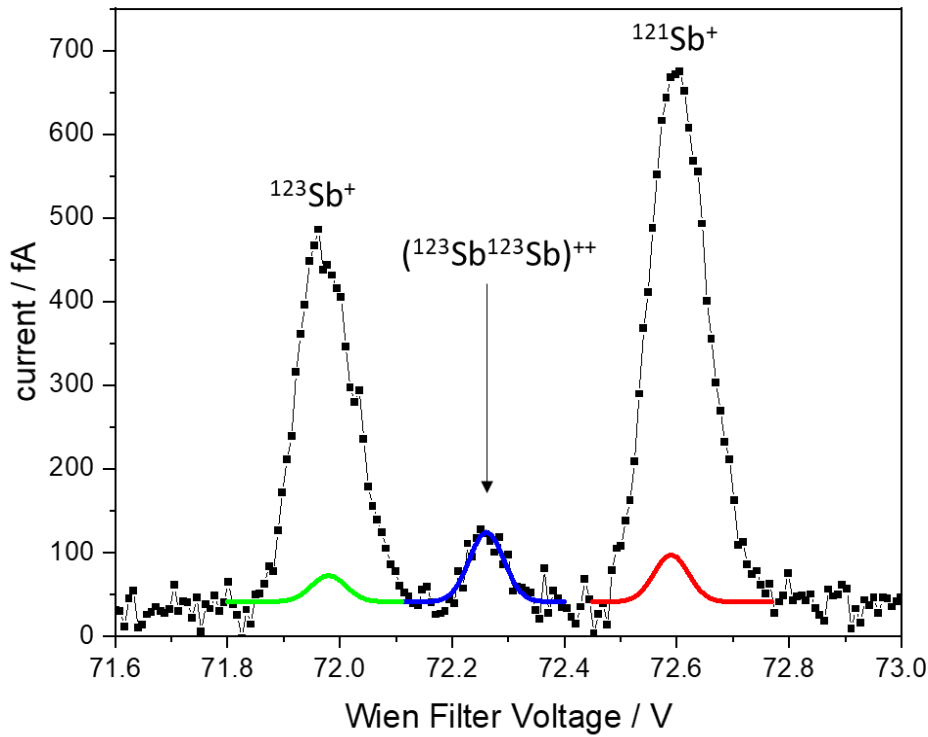

Figure 1- Wien filter (WF) scan of the AuSiSb liquid-metal-alloy source emission at 15 kV anode voltage, high mass resolution, low dose mode. Region of interest shows the region where singly-charged Sb and the $Sb_2$ clusters are found. The measured data (black squares) are fitted based upon the isotopic native abundance (coloured lines).

Confidence in the identification of the two Sb peaks comes from their position in the Wien filter scan and analysis of their relative areas, with the two peaks being 57.2% and 42.8% consistent with literature (57.21% and 42.79%) for $^{121}$Sb and $^{123}$Sb [24]. To confirm this a 5 μm x 5 μm region of an intrinsic Si wafer was implanted with singly charged Sb to a dose of 1E15 ions.cm$^{-2}$ at each Wien filter voltage associated with the three peaks observed in Figure 1. Figure 2(a) and (b) show NanoSIMS depth profiles for the $^{121}Sb^+$ and $^{123}Sb^+$ implanted regions, with only a single Sb isotope being found in each case. Figure 2(c) displays the NanoSIMS depth profiles for the $(^{121}Sb^{123}Sb)^{2+}$ implantation to the same dose as used in Figure 2(a) and (b), clearly showing the presence both Sb isotopes (each with a reduced relative concentration compared to the monoisotopic regions). All counts are normalised to $^{28}Si$ counts found in the $^{nat}Si$ wafer at long range. It is worth noting the NanoSIMS obtained depth profile of the implantation does not vary significantly between singly-charged single Sb ion and doubly-charged $Sb_2$ cluster implantation, indicating that the energy spread of the implanted ions, centred around 25 keV for each ion, is small as to be expected.

The Sb doping profile of 1E15 ions.cm$^{-2}$ at 25 keV into Si was simulated using TRIDYN, a dynamic Monte Carlo simulation for ion implantation [25], and is compared to the measured NanoSIMS depth profiles of $^{121}$Sb and $^{123}$Sb, in Figure 2(a). The simulation predicts a peak in the Sb doping concentration at a depth of 22 nm which is consistent with the experimentally measured peak Sb implantation depth of 18 nm, when considering the error associated with the ~10 nm surface accumulation layer necessary for steady state NanoSIMS analysis. However, the TRIDYN simulations predict a much faster decay of the implantation profile than is observed experimentally, indicating potential knock-on effects of the Sb. This knock-on implantation effect may result either from the NanoSIMS ion bombardment during characterisation or from the implantation high-dose rate used to achieve sufficient Sb concentrations for these tests. The latter will not affect use of Sb in qudit applications as Sb would be implanted as single clusters and the former is a measurement artefact, suggesting the TRIDYN simulations can be taken as a reliable predictor of the statistical distribution of implanted Sb clusters. It is noted that TRIDYN is not adapted to model concentrations below 0.1% (a normalised counts of ~3E-4 in Figure 2) meaning beyond a depth of 45 nm the predicted concentration should be treated with caution.

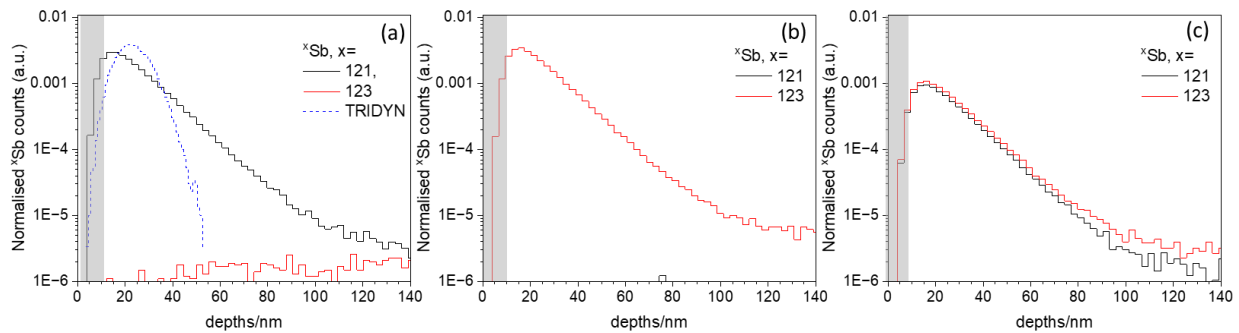

Figure 2 – NanoSIMS depth profiles of $^{121}$Sb and $^{123}$Sb, normalised to the bulk native $^{28}$Si count obtained from the sample. (a) $^{121}$Sb$^+$ implantation at 105 V (b) $^{123}$Sb$^+$ implantation at 106 V and (c) ($^{121}$Sb$^{123}$Sb)$^{2+}$ cluster implantation at 72.25 V. (a) also shows the simulated TRIDYN profile for Sb$^+$ 25keV 1E15 ions.cm$^{-2}$ implantation (blue dashed line). The depths were calibrated using post NanoSIMS atomic force microscopy. The grey shaded region at shallow depths indicates the pre-steady state NanoSIMS collection regime.

The NanoSIMS analysis in Figure 2 has demonstrated the controlled implantation of equal amounts of $^{121}$Sb and $^{123}$Sb using Sb$_2$ clusters, but the ~50 ppm detection limit of this approach is too low to allow measurement of the single Sb$_2$ clusters required for qudit fabrication. Aberration corrected STEM imaging is one of the only techniques able to detect single atomic clusters buried within the implanted film. To avoid the potential artefacts that could be created by cross sectional sample preparation, ($^{121}$Sb$^{123}$Sb)$^{2+}$ molecules were directly implanted into pre-prepared 20 nm thick Si membranes (see SI Figure S2) at an energy of 30 keV (15 kV anode voltage). TRYDIN simulations indicate that at this voltage ~76% of implanted Sb atoms will stop within the membrane. A pulsed ion current was set such that within each pulse one Sb$_2$ molecule would be expected based on Poisson statistics ($\lambda$ = 1). The FIB implantation 'spot-size' was 60 nm (determined using the edge method previously described [20]) and single pulses were implanted with a 15 nm step size with the aim of implanting isolated Sb$_2$ clusters suitable for STEM imaging.

Annular dark field (ADF) STEM imaging of the implanted membrane is shown in Figure 3(a), where the pairs of Sb ion dopants are highlighted via white rectangles. Implanted Sb ions are visible due to the higher atomic number of Sb compared to the surrounding Si lattice, with a local ~33% increase in the ADF intensity for a Sb substituted Si column measured experimentally compared to the surrounding lattice (Figure 3(c)). Larger magnitude, but longer period, intensity variations are also observed in the Si lattice, but these are due to thickness variations in the membrane as well as the presence of surface contamination. Multislice image simulations show that to optimise imaging conditions, a semi convergence angle of 21 mrad with ADF collection angles of 30-50 mrad (SI Figure S3) is needed. The predicted average HADF contrast of single Sb dopants in 20 nm thick Si is 34%, consistent with the experimental observations (see SI Figure S4). Imaging the nearest neighbour Sb-to-Sb distances for 176 dopant ions revealed a peak Sb-Sb distance of ~2 nm (Figure 3(d)) consistent with that expected for the implantation of ($^{121}$Sb$^{123}$Sb) clusters. A further 30 isolated ion dopants were identified where no other Sb peak was observed within the 1445 by 1445 nm field of view. For these isolated ions the pair from the cluster is expected to be either one of the 24% that are predicted to have passed through the film during implantation, or to be at a sufficiently different depth in the sample that due to the ~10nm depth of field of the STEM image, they are not visible in a single frame.

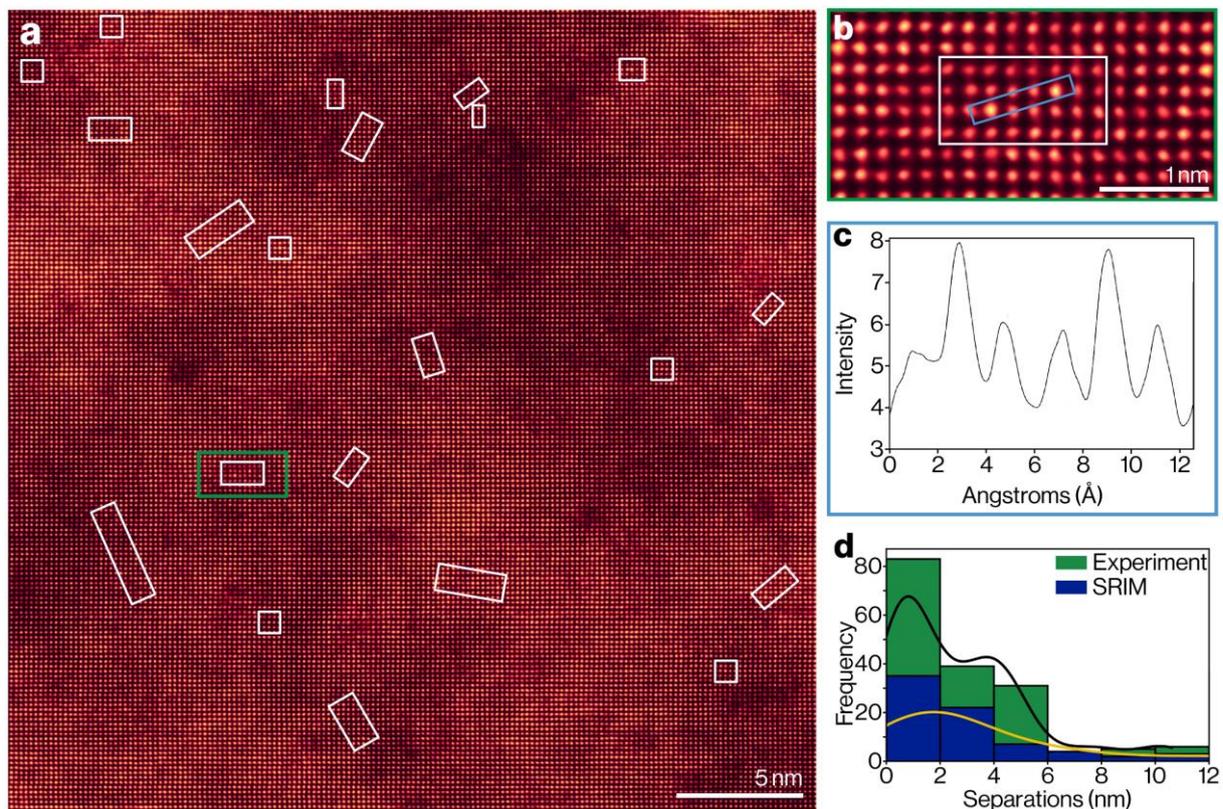

Figure 3 – ADF STEM imaging of Sb-implantation in Si. (a, b) ADF STEM images of Si membrane highlighting the presence of Sb ions resulting from the implanted single ($^{121}$Sb$^{123}$Sb)$^{2+}$ clusters with a planar spacing of 6.2 Å. Single- and double-ion and dopants are highlighted *via* white squares and rectangles, respectively. (c) Intensity line profile demonstrating the ~33% increase in peak intensity resulting from substitution of a single Sb ion into the Si lattice. (d) Histograms showing the experimentally measured Sb-to-Sb separation compared with the expected distribution predicted from SRIM Monte Carlo simulations [26] Kernel density estimates were fitted to the experimental and simulated histograms using Python's Seaborn library, shown via the black and yellow splines, respectively.

To be utilised as solid-state qubits, the isotopically selected Sb$_2$ clusters must be implanted into Si in a deterministic manner with high confidence (detection efficiency, η). High detection efficiency has already been reported with the P-NAME tool using Sb ions: 87 (±7)% for 50 keV Sb ions into SiO$_2$ [20]. However efficiencies vary significantly for different ion beam species, implantation energies, and target material combinations [27]. For the Sb in Si system considered here, measurements of η were undertaken using the method outlined by Cassidy, *et. al*. [27], more details of which are provided in the Experimental Section.

The implantation energies chosen for this investigation used the maximum anode voltage of the P-NAME system (25 kV), which results in a nominal ion beam energy of 25 keV for a singly-charged ion. However, due to the nature doubly-charged of the clusters discussed above, there is no way to separate the isobars of the singly charged individual Sb ions from the doubly-charged clusters ($^{121}Sb_2^{2+}$ and $^{123}Sb_2^{2+}$) using the Wien filter of the P-NAME system. As a result, only comparisons between doubly charged ion species (both mono-atomic and molecular) are made here. The implantation of the mixed-isotope $Sb_2$ cluster exhibits the highest detection efficiency of 94 (±8)% compared to 70 (±5)% for a single $^{123}Sb$ ion at the same implantation energy per atom (SI Table S1). When the implantation energy per atom is increased to 50 keV (implantation of a doubly-charged ion at 25 kV anode), the detection efficiency is 81 (±3)%.

The correlative relationship between ion kinetic energy and secondary electron emission for ion-substrate combinations has been well documented for a variety of ion-material systems [28-30]. As such, an increase in detection efficiency by increasing the anode voltage from 12.5 kV to 25 kV for the doubly-charged ion (increasing implantation energy from 25 keV to 50 keV) is to be expected. The increase of the secondary electron emission (indicated by an increased detection efficiency) for the mixed-cluster species is also in line with previous work [31, 32], particularly with Sb clusters [33], which demonstrated that an increase in cluster size leads to an increase in secondary electron emission, though with a sublinear correlation between cluster size and average secondary electron yield per atom within the cluster.

## **Conclusion**

This study has demonstrated the ability to isolate isotopically unique ($^{121}Sb^{123}Sb$) clusters using a liquid metal alloy ion source. NanoSIMS analysis demonstrates isotopic doping of $^{121}Sb^{123}Sb$ clusters into Si with the expected $^{121}Sb:^{123}Sb$ ratio using the P-NAME system. Atomically resolved imaging of implanted $^{121}Sb^{123}Sb$ clusters reveals the expected atomic distribution and separations consistent with theoretical calculations. The observed isotopically selected $^{121}Sb^{123}Sb$ clusters with mean separations of ~2 nm following implantation are ideally suited to form coupled qudit systems due to overlap of their respective electronic wavefunctions. The effective demonstration of this for Sb ions provides the possibility to explore higher dimensional Hilbert spaces in the 3-in-1 qubit regime, previously only demonstrated with P. Analysis of the detection efficiency of single cluster ($^{121}Sb^{123}Sb)^{2+}$ doping with P-NAME yields a value of $\eta$ = 94 (±8)% [20], showing that deterministic doping of single clusters with high confidence is feasible. Coupled with the ability of the same instrument to also provide <3 ppm $^{29}Si$ host regions for these to reside [17] we here demonstrate a scalable pathway for fabricating large arrays of qudits. These results open up a new route to optimised Si-based quantum computing platforms.

## **Methods**

Implantation was performed using a mass-selected Sb ion beam provided by the Platform for Nanoscale Advanced Materials Engineering (P-NAME, Q-One Ionoptika) tool whose details are reported in full elsewhere [20]. The in-column Wien filter was used to mass-select Sb ions with an optimum Mass Resolving Power (MRP, M /ΔM) of 296 ± 7 for the $Sb^+$ isotopes under mid-current (in the range of 10's pA) mode at a 25 kV anode voltage. Implanted substrates were placed with a 3° offset with respect to the beam in order to minimise ion channelling.

The source alloy had $Au_{73}Si_{14}Sb_{13}$ near-eutectic composition produced by arc melting 99.9% gold ingot (Cookson Precious Metals Limited) with 99.99% Si and 99.5% Sb lump (both Thermo Fisher Scientific). Arc melting was conducted in 30 mTorr vacuum, backfilled with Ar on a Cu hearth. The liquid alloy was flipped three times, and then the power was stopped and the alloy solidified on the water-cooled Cu plate.

NanoSIMS analysis was performed using a NanoSIMS 50L (CAMECA, France), employing a 16 keV $Cs^+$ primary ion beam with a beam current of 0.4 - 0.5 pA and approximate beam size of 75-100 nm. It was scanned across the sample surface to generate negative secondary ions, with secondary electron (SE) images also collected. The instrument was tuned to achieve a MRP exceeding 6000 for all detectors, effectively mitigating any mass interferences. A double-focusing mass spectrometer facilitated the simultaneous detection of key ions $^{12}C$, $^{16}O$, $^{28}Si$, $^{31}P$, $^{121}Sb$, $^{123}Sb$ and SE. Imaging was conducted over 500 planes, each with a 256 × 256 pixel resolution and a 5 µm raster size, with a dwell time of 1000 µs per pixel. To enhance spatial resolution, the D1 aperture is adjusted to D1-4 (150 µm diameter) and entrance slits to ES-3 (30 µm width), alongside an aperture slit set to AS-2 (200 µm width). The NanoSIMS signals in Figure 2 have been normalised to the native $^{28}Si$ signal taken from a depth where there is no further Sb signal observed. It is observed that the $^{123}Sb$ signal is reproducibly higher across all samples, by about 13% which is accounted for by a systematic instrumental effect relating to the exact location of the collected mass. This was intentionally introduced to reduce mass interference in the NanoSIMS measurements and does not affect the results obtained. Atomic force microscopy (AFM) profiling was conducted using a nanoSurf CoreAFM with a tapAL-190 probe and was used the calibrating the NanoSIMS etch profiles.

Implantation for STEM was carried out into 20 nm thick Si membranes (Silson) using a 15 kV anode voltage, 30 keV implantation energy, Sb cluster beam. STEM characterisation used a Thermo Fisher Titan equipped with a Schottky field emission gun (FEG). STEM images were acquired using a ThermoFisher Titan ChemiSTEM at an accelerating voltage of 200 kV with a probe semi-convergence angle of 21 mrad, a probe current of 78 pA. Annular dark field (ADF) imaging was performed with detector inner/outer angles of 54/200 mrad and an electron fluence of 1950 $e^-Å^2$. Further STEM images, were acquired using a ThermoFisher Iliad STEM with an accelerating voltage of 300 kV, a probe semi-convergence angle of 21.4 mrad, a probe current of 14 pA. ADF imaging used inner/outer collection angles of 33-200 mrad and an electron fluence of 783 $e^-Å^2$ per image, requiring a relatively high signal-to-noise (S/N) to confidently resolve the implanted species with respect to the silicon. Nonetheless, it is desirable to limit the electron fluence since the high energy electron beam has the potential to cause displacement of the single-ion dopants. Immediately prior to STEM imaging the implanted membrane samples were cleaned using toluene at 50 °C to remove any volatile organic compounds adsorbed on the surface. Mutlislice simulations were performed with abTEM [34] using the Kirkland parametrisation [35-37] and PRISM algorithm [38]. For further information, see SI Section 2.

TRIDYN simulations were performed using the tridyn2020l version of the code. The input files were modified as described in the text, ensuring that each simulation achieved a maximum relative areal density change per pseudoprojectile of <1%, as required to ensure statistical validity of the simulation.

The detection efficiency measurements were conducted using the P-NAME internal high sensitivity secondary electron detectors. The ion beam currents were 105 (±4) fA, 86 (±6) fA, and 86 (±7) fA for the 50 keV monoatomic ion, 25 keV monoatomic ion, and 50 keV mixed cluster ion, respectively. The electrostatic pulser voltage (nominally high to keep the beam blank) was dropped to 0 V for a desired set time (the pulse width) such that the average number of ions per pulse is Poissonian in nature. The pulse width was varied for the average number of ions per pulse from 0.1 to 1, in 0.1 steps. When the beam was pulsed, the ion implanted into the substrate releases secondary electrons into the chamber, which are collected by the secondary electron detectors. The efficiency of this collection is dominated by the secondary electron emission, which is ion beam and material dependent.

In order to calculate the efficiency, the number of pulses required to achieve a total of 2500 successful implantation detection (secondary electron signals correlating with the opening of the electrostatic pulser) was recorded, and linear regression applied to the data [27] to determine the detection efficiencies for the different ion species. The ions were implanted in a 50 x 50 array with a pitch of 1 um between each implant, and each of the variations in pulser duration and ion beam species was implanted into a new region of the Si to avoid any potential effect of localised charging or secondary electron depletion. The dak count rate (false positive detections) of the P-NAME system was measured to be $k = 10^{-5}$ $s^{-1}$, and so not considered to have a significant impact on the measurements made in this work.

**Acknowledgements**

The authors acknowledge funding from the Engineering and Physical Sciences Research Council (EPSRC) for funding under grants EP/Y024303, EP/S021531/1, EP/M010619/1, EP/V007033/1, EP/S030719/1, EP/V001914/1, EP/V036343/1 and EP/P009050/1, and also for the EPSRC Centre for Doctoral Training (CDT) Graphene-NOWNANO. TEM and NanoSIMS access was supported by the Henry Royce Institute for Advanced Materials, funded through EPSRC grants EP/R00661X/1, EP/S019367/1, EP/P025021/1 and EP/P025498/1. The NanoSIMS was funded by UK Research Partnership Investment Funding (UKRPIF) Manchester RPIF Round 2. SJH and ET acknowledge funding from the European Research Council (ERC) under the European Union's Horizon 2020 research and innovation programme (Grant ERC-2016-STG-EvoluTEM-715502) and thank Chris Livingston (NGI, University of Manchester) for his support posting and receiving TEM specimens.